\documentclass[12pt,preprint]{aastex}
\usepackage{psfig}
\def\HI {H\kern0.1em{\sc i}} 
\def\deg{$^{\circ}$}

\begin{document}
\title{~~\\ ~~\\ Identifying Compact Symmetric Objects in the Southern Sky}
\shorttitle{Southern COINS}
\shortauthors{Taylor \& Peck}
\author{G. B. Taylor\altaffilmark{1} \& A.B. Peck\altaffilmark{2}}
\email{gtaylor@nrao.edu; apeck@sma.edu}
\altaffiltext{1}{National Radio Astronomy Observatory, P.O. Box 0, Socorro, NM 87801}
\altaffiltext{2}{Harvard-Smithsonian CfA,SAO/SMA Project, P.O. Box
  824, Hilo, HI 96720}


\slugcomment{As Accepted by the Astrophysical Journal}

\begin{abstract}

We present results of multifrequency polarimetric VLBA observations of
20 compact radio sources.  The observations represent the northern
and southern extensions of a large survey undertaken to identify
Compact Symmetric Objects (CSOs) Observed in the Northern Sky (COINS).
CSOs are young radio galaxies whose jet axes lie close to the plane of
the sky, and whose appearance is therefore not dominated by
relativistic beaming effects.  The small linear sizes of CSOs make
them valuable for studies of both the evolution of radio galaxies and
testing unified schemes for active galactic nuclei (AGN).  In this
paper we report on observations made of 20 new CSO candidates
discovered in the northern and southern extremities of the VLBA
Calibrator Survey.  We identify 4 new CSOs, and discard 12 core-jet
sources.  The remaining 4 sources remain candidates pending further
investigation. We present continuum images at 5 GHz and 15 GHz
and, where relevant, images of the polarized flux density and spectral
index distributions for the 8 new CSOs and CSO candidates.

\end{abstract}


\section{Introduction}

Compact Symmetric Objects (CSOs) are a recently identified class of
radio source which are less than 1 kpc in size, and are thought to be
very young objects ($\le$10$^4$~yr, \cite{rea96a, ows98a}).  These
sources are uniquely well suited to investigations into the physics of
the central engines and the evolution of radio galaxies.  CSOs have
proved useful for testing the unified scheme of active galactic
nuclei, which requires an obscuring region of atomic or molecular gas
surrounding the central engine and effectively shielding the inner few
parsecs of the source from view if the radio axis lies at a large
angle to the line of sight \citep{ant93}.  This model is supported by
observations of a few symmetric radio sources which exhibit very
broad H\kern0.1em{I} absorption lines \citep{tay99, pec01}, and 
free-free absorption ({\it e.~g.} \citet{pec99}).  It is 
reasonable to assume that if CSOs are young, this
circumnuclear material is accreting onto, and ``feeding'', the central
engine, and that this process will lead to their eventual evolution
into much larger FR II sources \citep{rea96b,fan95}.

Measurements of the hotspot advance speeds can provide a kinematic
age estimate for CSOs.  To date, hotspot advance speeds have been reported
for fewer than 10 sources (see \citet{pol03} for a review).
The velocities reported are all of order 0.2 $h^{-1}$ c
(where $h$ = H$_0$/100 km s$^{-1}$ Mpc$^{-1}$),
yielding kinematic age estimates of between 200 and 2000 years.

Considerably larger velocities have been seen for the jet components
of CSOs, and may provide a means of constraining H$_0$.  The symmetry
axes of CSOs are known to be at moderate to large angles, $\theta$, to
the line-of-sight \citep{rea96a}, such that counterjet components are
frequently visible.  If the jets have bulk velocities, $\beta$, then
the separation velocity between oppositely directed components,
together with their velocity ratios, distance ratios, and flux density
ratios, constrain the allowed values for $\beta$, $\theta$, and
H$_0$ \citep{tay97}.  Once the velocities of jet/counterjet pairs in
enough CSOs are measured, then it may also be possible to constrain
cosmological parameters since CSOs are found at a wide range of redshifts.  

In addition to their scientific appeal, CSOs can also be quite 
useful calibrators.  CSOs have been shown
to be remarkably stable in flux density \citep{fas01}, making them
ideal sources to use as amplitude calibrators in monitoring
experiments, such as measuring the time delay between components of
gravitational lens systems.  CSOs to date have been observed to
have very low fractional polarization \citep{pec00}, making them
useful to solve for leakage terms, or as an independent check
on the quality of the polarization calibration.

Here we report on efforts to identify CSOs in the Northern cap (dec
$>$ 75\deg) and in the Southern sky ($0$\deg\ $>$ dec $>$ $-$30\deg).  This will
complete our survey to identify valuable sources (CSOs in the Northern
Sky (COINS), \citet{pec00}) which can be searched for
H\kern0.1em{I} and free-free absorption, used as calibrators, and are
promising candidates to be added to our sample for proper motion
studies.

We assume H$_0 = 70$ km s$^{-1}$ Mpc$^{-1}$,  $\Omega_M$=0.27, and a
flat cosmology throughout.
  
\section{Sample Selection}

We have been working on identifying a substantial sample of CSOs 
starting from a number of VLBI surveys including the Pearson-Readhead 
(1988) \nocite{pr88} sample, Caltech-Jodrell Surveys 
(CJ1 -- \citet{pol95}; CJ2 -- \citet{tay94}; CJF - \citet{tay96}), and the 
VLBA Calibrator Survey (VCS - \citet{bea02}). The initial sample
concentrated on CSOs Observed in the Northern Sky (COINS).  With the
extension of the VCS to southern declinations (0\deg\ to $-$30\deg) 
and also the northern cap (dec $>$ 75\deg) we have been able to 
extend the COINS sample.  

Some 425 sources were examined from the VCS at both the 8.4 and 2.2
GHz bands.  Sources were selected according to the criteria used by
\citet{pec00}.  Namely these are objects having (1) a nearly equal
double structure (intensity ratio less than 10) at either 2.2 or 8.4
GHz or (2) a strong central component with extended emission on both
sides at one or both frequencies.  Sources exhibiting
edge-brightening of one or more components were given priority.
These selection criteria are not among the defining characteristics
of CSO's, but are used to eliminate the more obvious ``core-jet''
sources which comprise the majority of sources in this flat spectrum
survey.   These selection criteria applied to the northern and
southern extremities of the VCS yielded 20 CSO candidates for further
study.

\section{Observations and Analysis}

The observations, performed on 2002 April 26, were carried out at 5
and 15.2 GHz using the VLBA\footnote {The National Radio Astronomy
Observatory is operated by Associated Universities, Inc., under
cooperative agreement with the National Science Foundation.}. Due to
problems with the site computer, the VLBA element at North Liberty did
not observe, thus reducing the VLBA to 9 stations for this experiment.
Amplitude calibration was performed in the standard way using
measurements of the system temperatures and antenna gains.
Fringe-fitting was performed on the strong calibrator 3C279 and a
moderately strong target source, J0240-2309.  The delays determined
from J0240$-$2309 were used for all target sources.  Feed
polarizations of the antennas were determined at 5 GHz only using the
AIPS task LPCAL. This calibration was performed twice, first using
J0240$-$2309 with an accompanying CLEAN model for this polarized
source. After this calibration was performed J1935$+$8130 was found to
be unpolarized, and although somewhat weaker, it had the benefit of
better parallactic angle coverage. The D-terms were solved for again
using J1935$+$8130 as an unpolarized calibrator. Both sources produced
nearly identical corrections for the D-terms, and J1935$+$8130 was
used as the final D-term calibrator.  No polarimetry was attempted
for the 15 GHz observations, primarily because the sources are all
considerably weaker at this frequency.  

Absolute electric vector position angle (EVPA) calibration was
determined by using the EVPA's of 3C279
listed in the VLA Monitoring
Program\footnote{http://www.aoc.nrao.edu/$\sim$smyers/calibration/}
\citep{tmy00}.  A total of 1.15 Jy of polarized flux was seen
in the VLBA observations compared to 1.11 Jy seen by the VLA 
monitoring observation on 25 April 2002.  For the current experiment the 
absolute EVPA is of secondary importance to detecting, or placing
limits on, the linearly polarized flux density.  No attempt has been
made to correct the electric vector polarization angles for 
Faraday rotation, which is often significant on the parsec scale
for AGN \citep{zav03}.

For each source we tapered the 15 GHz data to produce an image at
comparable resolution to the full resolution 5 GHz image.  We then
combined the two images to generate a spectral index image.  It is
important to note that spectral index maps made from two datasets
with substantially different ($u,v$) coverages (such as we present
here) may suffer from significant systematic errors, especially in
regions of extended emission.

\section{Identified CSOs}

In the following section we display results confirming 4 sources as
bona-fide CSOs from our initial 20 candidates identified from the
extremities of the VCS. We also present 4 sources we still regard as
CSO candidates.  CSOs are defined by linear size and symmetry rather
than assumed youth of the object based on spectrum and compactness.
Thus it is important to remember that not all
gigahertz-peaked-spectrum (GPS) sources are CSOs.  Ideally, a core
component with a flatter spectrum than the hotspots on either side
must be identified before sources can be confirmed as CSOs.  For some
CSOs with a jet axis very close to the plane of the sky the core may
be so weak as to be undetectable yet a CSO identification can still be
secured if there are symmetric edge-brightened hot spots
and/or extended lobes.  On the basis of morphology and spectral index
(defined $S_{\nu} \propto \nu^\alpha$), we reject 12 sources as
core-jets.  Although not used as a defining characteristic, the
polarimetry results for the sources identified as CSOs are consistent
with previous observations of CSOs.  Identifications and redshifts are
given for each target source in Table 1.  In Table 2 properties of the
images are given, along with the current classification of each source
as CSO or Candidate CSO.  And in Table 3 we present estimates of the
properties (flux densities, spectral index, and polarized intensity)
of dominant components within each source.

\subsection{\bf J0242{\tt -}2132 (Candidate)}  This source has been identified
with a type ``N'' galaxy at a redshift of 0.314 \citep{wri90}.  The 5
GHz VLBA image (Fig.~1) shows a compact feature (labeled 'B') with
extended emission on either side.  While it is tempting to identify
component 'B' with the center of activity, this component has a steep
spectral index of $-$0.67 $\pm$ 0.05. The 5 GHz polarimetry indicates
that component 'B' is polarized at the 1-3\% level, with the maximum
in polarization occurring on the eastern edge of the source. Generally
the cores of radio galaxies are much less polarized \citep{ptz03}.
Some flat-spectrum ($\alpha \sim 0.29$) emission is seen to the
southwest of the peak, but there is no compact feature that can be
unambiguously identified as the core.  We conclude that this source is
a good CSO candidate as the core is unlikely to be located at either
extremity, but the exact nature of the source remains a mystery.  If
it is confirmed as a CSO it will be the first to be detected with
polarized flux.


\subsection{\bf J0425{\tt -}1612 (CSO)} Very little is known about this
radio source at optical wavelengths.  At 5 GHz (Fig.~2a) the source
consists of a bright, slightly extended component to the southwest
('D').  It then extends in a linear series of components to the
northeast extending over 80 mas, before bending by 90 degrees
(component 'A').  A spectral index image (Fig.~2b) shows that the 
three strongest components are all fairly steep
(between $-$0.6 and $-$1.0).  There is a faint, compact, and inverted
spectrum ($\alpha = 0.2 \pm 0.4$) component ('C') that we identify 
with the center of activity.  There is no detected polarized flux
stronger than 0.72 mJy.  We confirm this source as a CSO.


\subsection{\bf J0735{\tt -}1735 (CSO)} This radio source has been
used as a calibrator for the Radio Reference Frame \citep{joh95}.
At 5 GHz (Fig.~3a) the source appears to have
the shape of a boomerang with the brightest component in the center.
The entire source is steep spectrum with the center component
marginally flatter ($\alpha = -0.54$) than the edge components.  In the full
resolution 15 GHz image (Fig.~3c), we see a compact component 
at the base of a predominantly one-sided jet.  Some extended 
emission is visible on the counterjet side.  We identify the
compact component at the base of the jet with the center of 
activity.  The core appears unpolarized, but there is a marginal
detection of polarization at the level of 0.2\% in the unresolved
northern jet seen at 5 GHz.  The polarization at the edge of
the source reaches nearly 10\% but given the noise in the image
this is unlikely to be significant.  We confirm this source as a CSO.


\subsection{\bf J1211{\tt -}1926 (CSO)}  No optical counterpart has
been identified for this source.  At 5 GHz (Fig.~4) the source has a
classic CSO morphology with a compact inverted-spectrum core flanked
by extended, steep-spectrum emission.  The spectral index of component
A on the east side ($-$0.59) is marginally steeper than that of
component D on the west end ($-$0.53).  No polarized flux density is
detected to a limit of $<$1 mJy.  We confirm this source as a CSO.


\subsection{\bf J1248{\tt -}1959 (Candidate) } This source is identified with 
a 20.5 magnitude quasar at a redshift of 1.275 \citep{ode91}.
Both the 5 and 15 GHz images have
rather high noise owing to the orientation of the source combined with
the poor ($u,v$) coverage obtained.  The fidelity of the images is
also probably quite low.  In Fig.~5a. we identify two steep-spectrum
components embedded and connected by diffuse emission.  The more
compact of these two components, B, has a marginally flatter spectrum
($-$0.74 vs $-$0.92).  A possibly inverted spectrum at the north end
of the source is the result of essentially no emission at 5 GHz.
Further observations are needed to confirm the existence of this
feature.  There is less than 1.5 mJy of polarized flux.  

Based on the rather inconclusive data presented here, we cannot 
exclude the possibility that this source is a CSO,
though that seems unlikely.  Further observations will be 
needed to identify the location of the core.


\subsection{\bf J1419{\tt -}1928 (Candidate)}  This source is associated 
with a 
magnitude 17.5 Seyfert 1 host galaxy at a redshift of z=0.12000+/-0.0005
\citep{hun78}.  At 5 GHz (Fig.~6a), the source consists of a few roughly
colinear components.  Since we only obtained 4 scans, the sensitivity
and ($u$,$v$)
coverage is relatively poor.  Some polarized flux is detected from
the stronger components.   At 15 GHz (Fig.~6b) the full resolution image
shown in Fig.~6b resolves the source into a narrow jet.  The
location of the core component is not known.  The spectral index
is poorly determined due to the low quality of the 5 GHz data, 
but overall the source has a steep spectrum ($\alpha = -0.3$).  This
source remains a CSO candidate.


\subsection{\bf J1935{\tt +}8130 (Candidate)} This source, 
identified with a magnitude 21.1 galaxy, is the only candidate found
among the northern cap sources.  Fig.~7a shows a fairly
classical CSO structure with two strong components connected by some
fainter emission.  The eastern component, A, is approximately twice as
bright as the western one, C, and has a flatter spectrum ($-$0.66 vs
$-$0.94).  Given this difference between the spectra it is possible
that component A is the core.  No polarized flux stronger than 1.5 mJy
was detected.  This source remains a CSO candidate pending further
observations.


\subsection{\bf J2347{\tt -}1856 (CSO)}  
Yet another little-studied source, there is a very faint object on
the DSS, but as yet no identification or redshift. Fig.~8a shows a
double radio source with components of very nearly equal flux density
and spectra.  No polarized flux stronger than 1.5 mJy was detected.
Both components exhibit some faint extensions to the
west, and are resolved at 15 GHz, but unresolved at 5 GHz.
This structure is reminiscent of 1031+567 \citep{tay00}, a well-studied
CSO which is also lacking a detectable core.
Given the extended, edge-brightened character of the lobes in J2347{\tt -}1856, it seems
reasonable to assume that a weak core must be located somewhere
between them.  We confirm this source as a CSO.

\section{Current Status of the COINS Survey}

Starting from VCS images of 425 sources we have selected 20 sources
for followup VLBA observations at 5 and 15 GHz.  From these, we have
identified 4 CSOs and 4 CSO candidates.  This detection rate of 2\% is
quite similar to that obtained for the northern part of the VCS
\citep{pec00}.  As pointed out by \citet{pec00}, this detection rate
is much lower than that found in the PR sample of 7/65 (11\%), or
18/411 (4.4\%) found in the combined PR and CJ samples.  We expect a
slightly lower detection rate because the parent sample is comprised
predominantly of flat spectrum sources.  Nonetheless, this
significantly lower detection rate may be a result of the reduced
sensitivity and ($u,v$) coverage compared to the PR and CJ surveys, or
it may be a real effect of looking at the fainter end of the
luminosity function.  To eliminate the possibility of missed sources
due to observational restrictions, will require a large survey with
observations of comparable quality to the follow-up study presented
here.  Simulations might also be useful, but would be hampered by the
wide range of morphologies possible for CSOs.

Additional followup observations of CSOs and candidates identified in
\citep{pec00} are currently ongoing and will be presented, along with
a comprehensive list of all CSOs discovered in the COINS survey to
date, in a future paper. Deep 15 GHz observations, as well as
spectral line VLBI observations to search for neutral hydrogen and
polarization studies, can help to confirm or reject sources still
classified as CSO candidates.  Proper motion studies to pinpoint the
center of activity are also underway.  Additional redshift data and
flux densities at a wider range of frequencies will be presented as
well.  At that time, it will be possible to undertake statistical
analyses of many CSO properties based on our complete sample,
including range of redshifts and linear sizes, polarization levels,
and flux densities.

\acknowledgments 
We thank the referee for several insightful comments.  
This research has made use of the NASA/IPAC
Extragalactic Database (NED) which is operated by the Jet Propulsion
Laboratory, Caltech, under contract with NASA.

\clearpage

\begin{figure}
\vspace{20.4cm}
\includegraphics{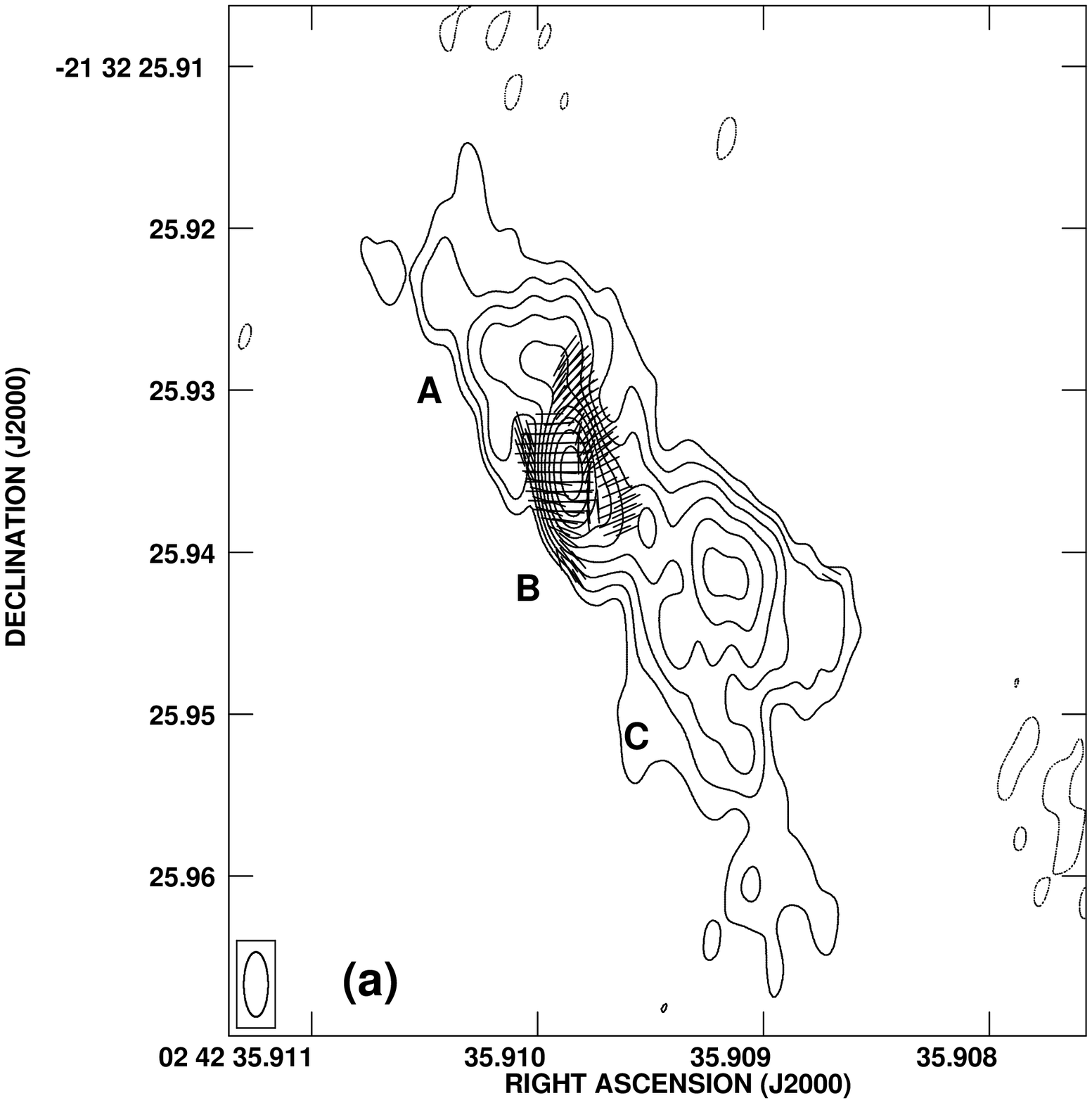}
\includegraphics{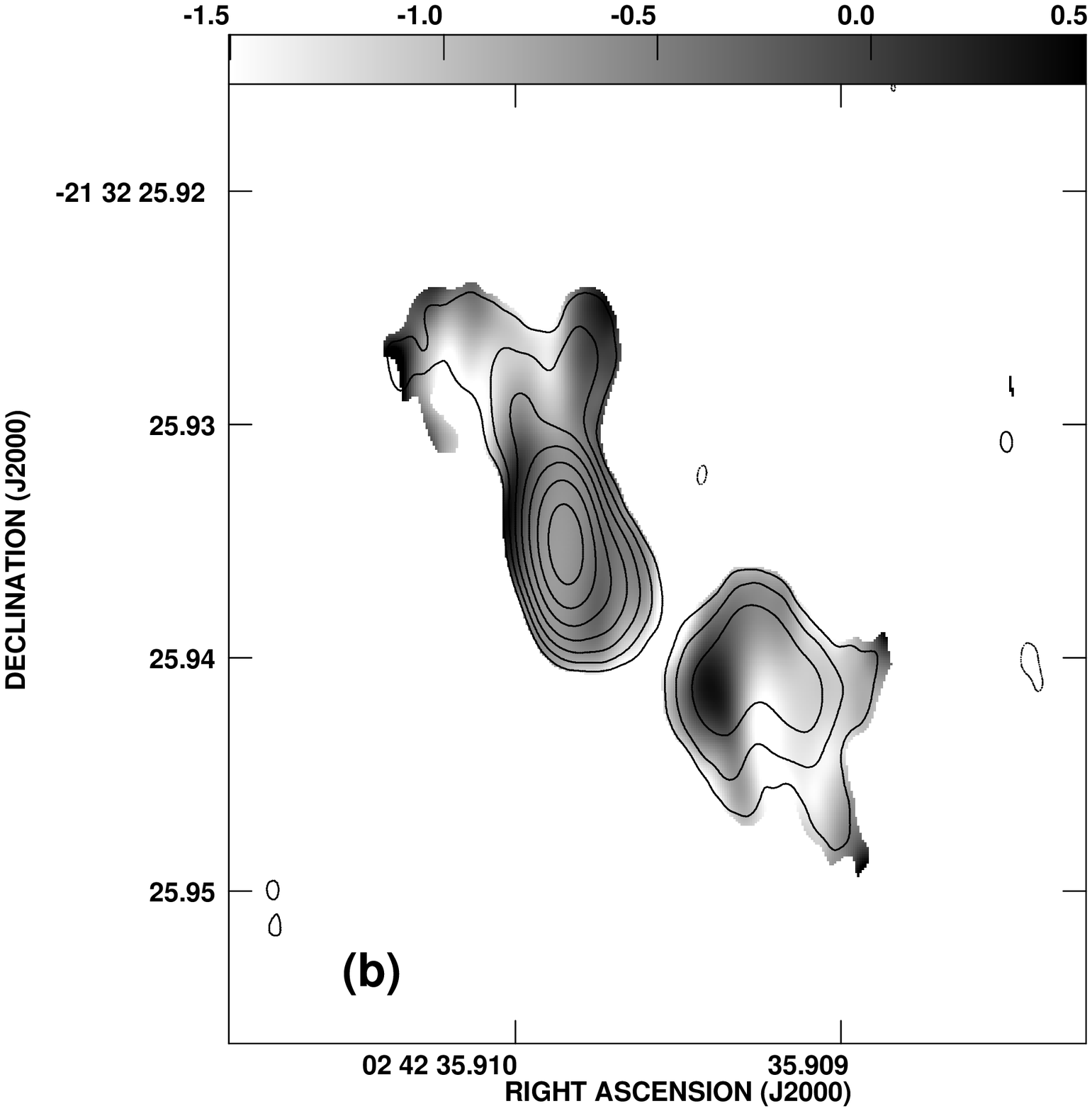}
\caption{{\bf (a)} Total intensity contours of J0242{\tt -}2132 at 5 GHz 
with electric 
polarization vectors overlaid.  A vector length of 1 mas corresponds
to a polarized flux density of 0.67 mJy/beam.  {\bf (b)} Total intensity
contours from 15 GHz with a greyscale map of the spectral index
between 5 and 15 GHz overlaid.  The greyscale range is from 
$-$1.5 to 0.5.  The restoring beam for both images is 4 $\times$ 1.25 
mas in position angle 0\deg.
\label{fig1}}
\end{figure}
\clearpage

\begin{figure}
\vspace{20.4cm}
\includegraphics{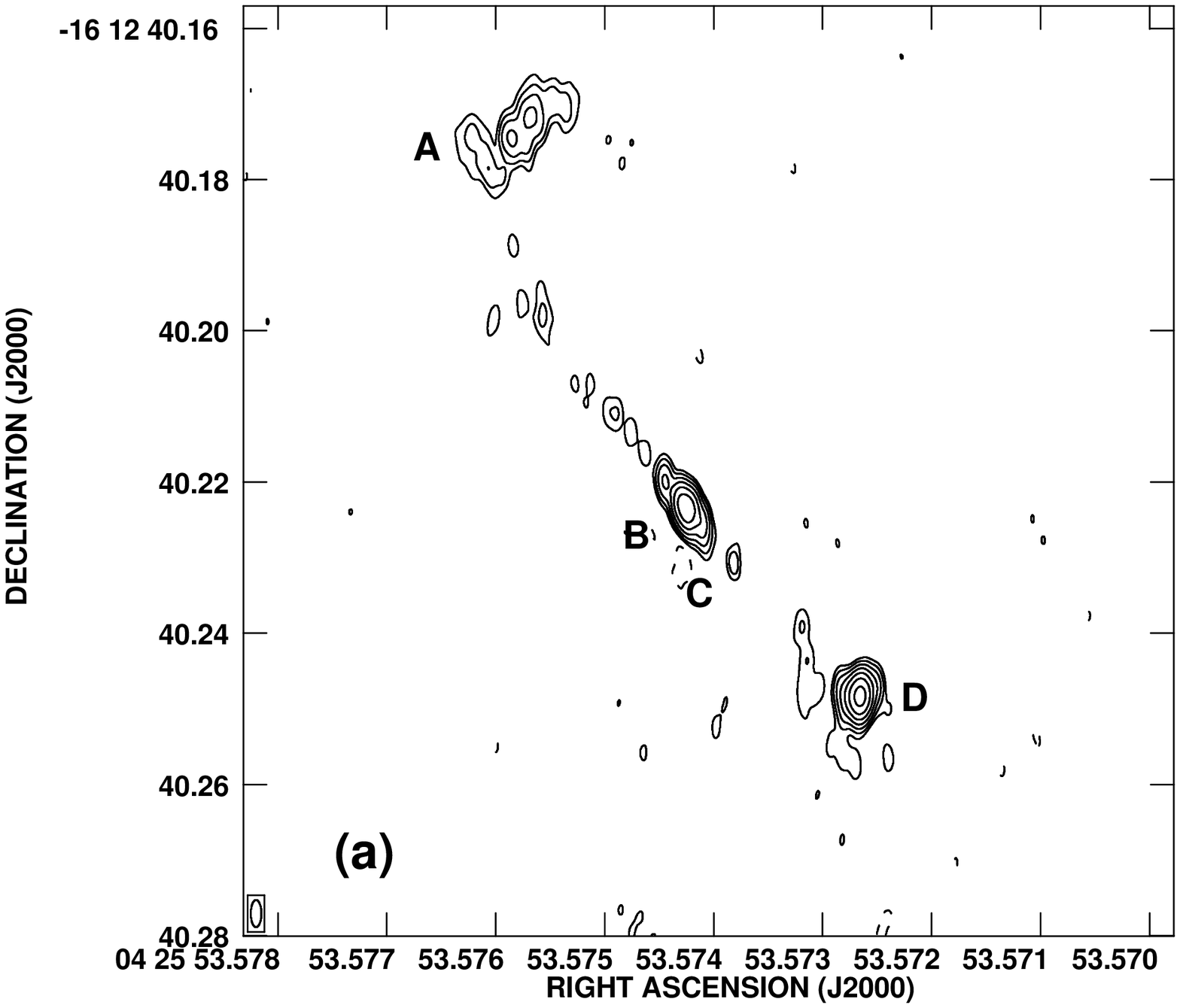}
\includegraphics{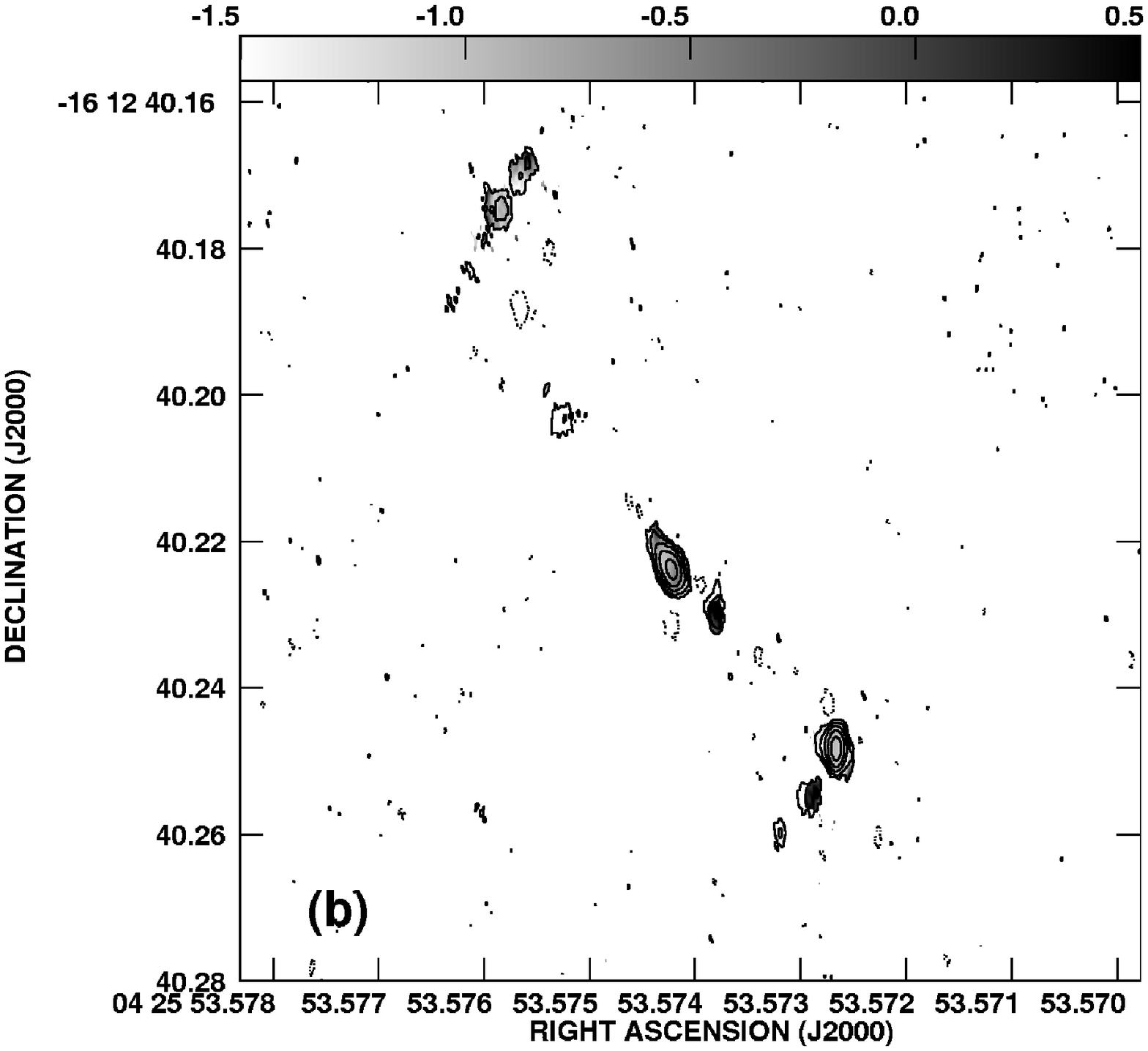}
\caption{{\bf (a)} Total intensity contours of J0425{\tt -}1612 at 5 GHz. 
{\bf (b)} Total intensity
contours from 15 GHz with a greyscale map of the spectral index
between 5 and 15 GHz overlaid.  The greyscale range is from 
$-$1.5 to 0.5.  The restoring beam for both images is 3.6 $\times$ 1.4
mas in position angle 0\deg.
\label{fig2}}
\end{figure}
\clearpage

\begin{figure}
\vspace{20.4cm}
\includegraphics{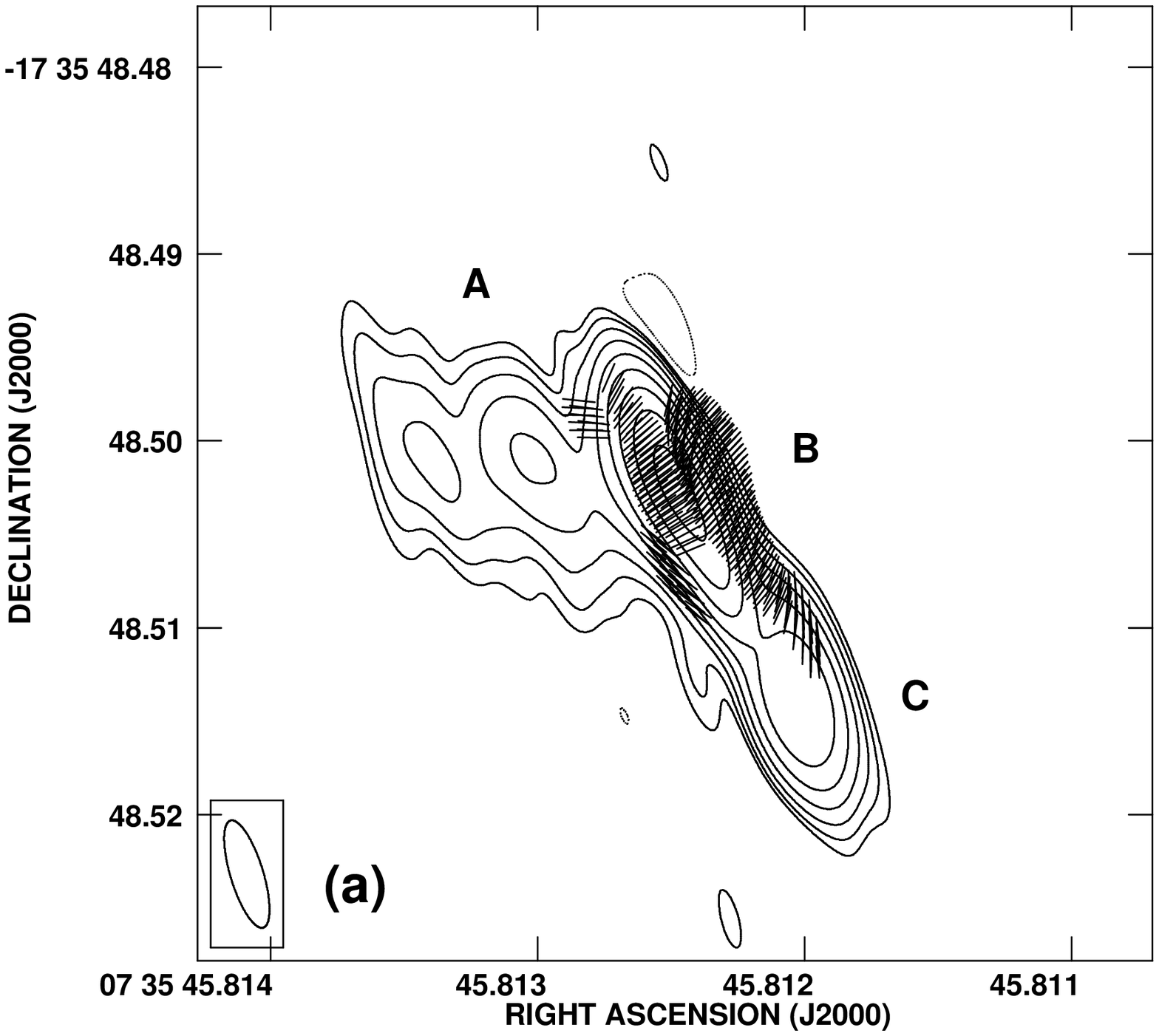}
\includegraphics{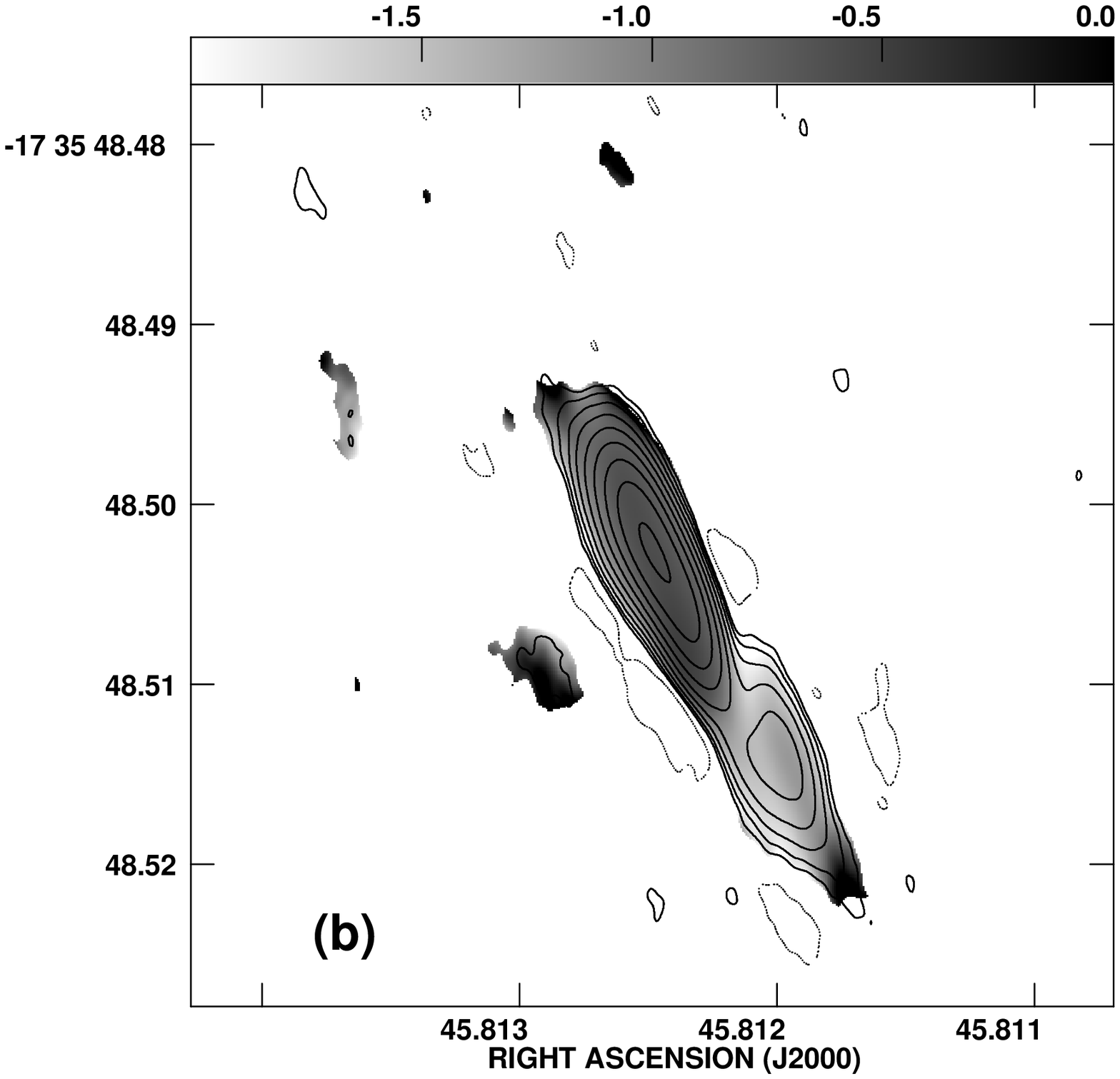}
\includegraphics{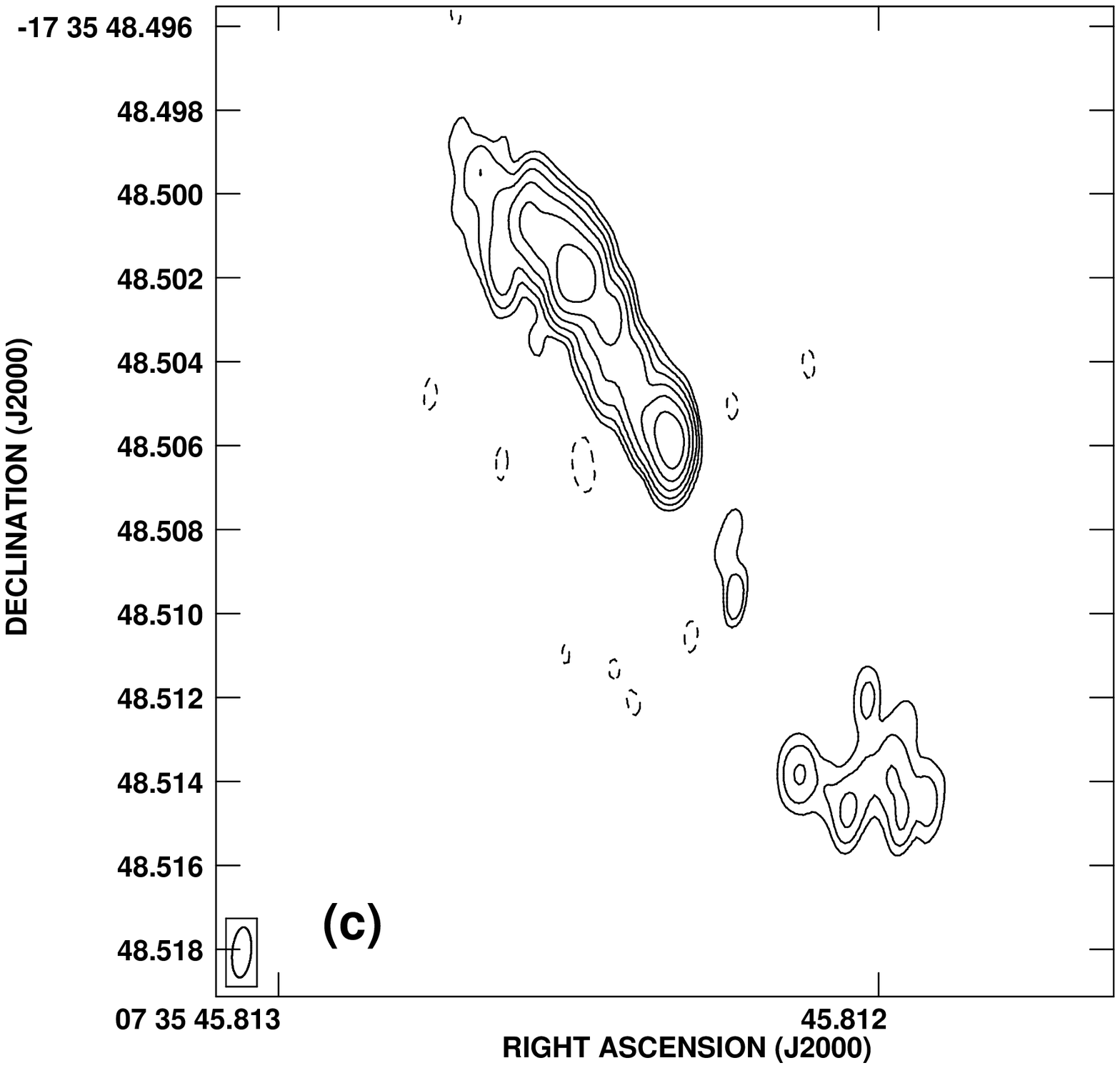}
\caption{{\bf (a)} Total intensity contours of J0735{\tt -}1735 at 5 GHz with 
electric 
polarization vectors overlaid.  A vector length of 1 mas corresponds
to a polarized flux density of 1 mJy/beam. 
{\bf (b)} Total intensity
contours from 15 GHz with a greyscale map of the spectral index
between 5 and 15 GHz overlaid.  The greyscale range is from 
$-$2.0 to 0.  The restoring beam for 3a and 3b is 6 $\times$ 1.75 
mas in position angle 17\deg.
{\bf (c)} Total intensity contours from the full resolution 15 GHz image.
The restoring beam is 1.2 $\times$ 0.45 mas in position angle -5.3\deg.
\label{fig3}}
\end{figure}
\clearpage

\begin{figure}
\vspace{20.4cm}
\includegraphics{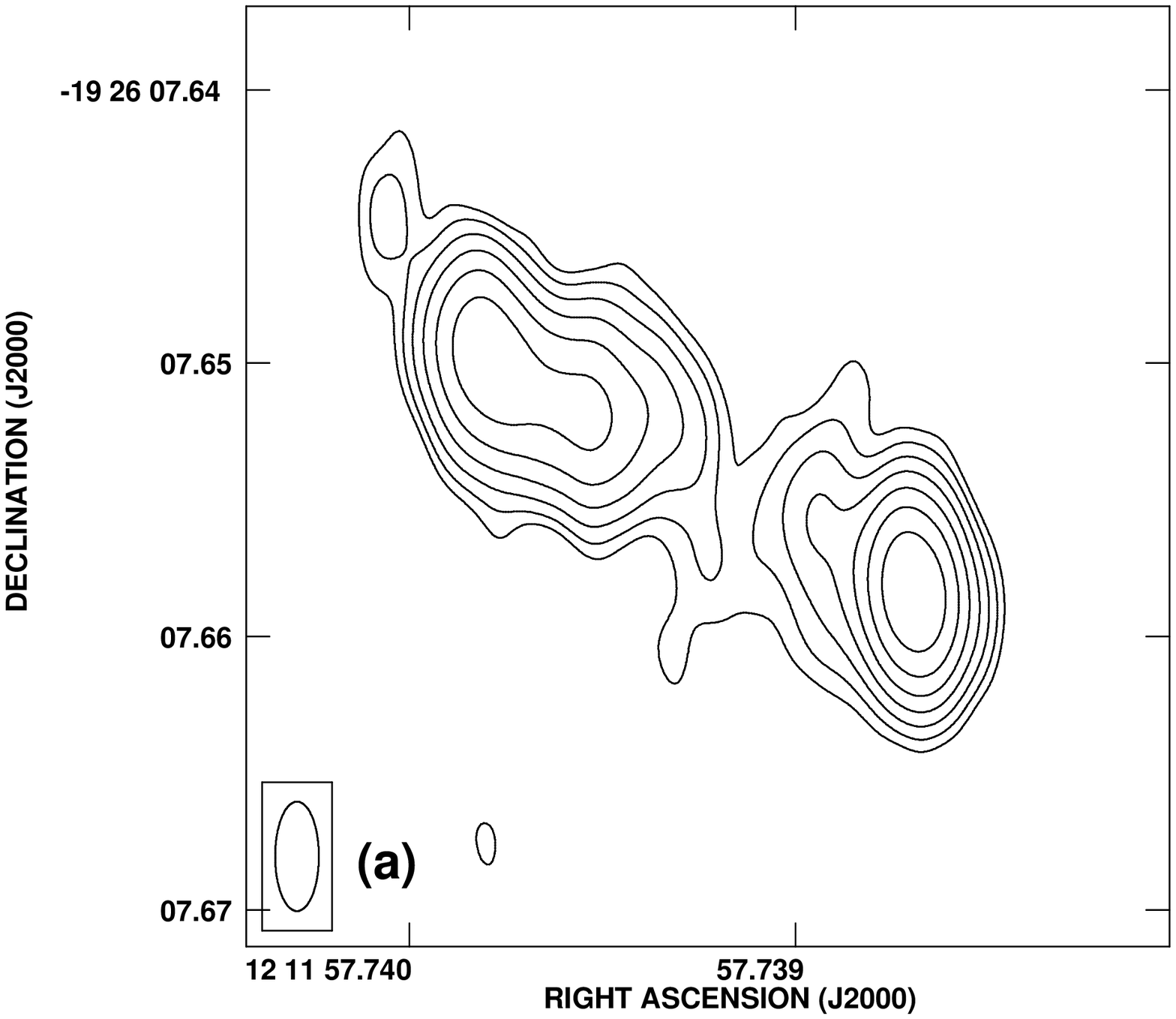}
\includegraphics{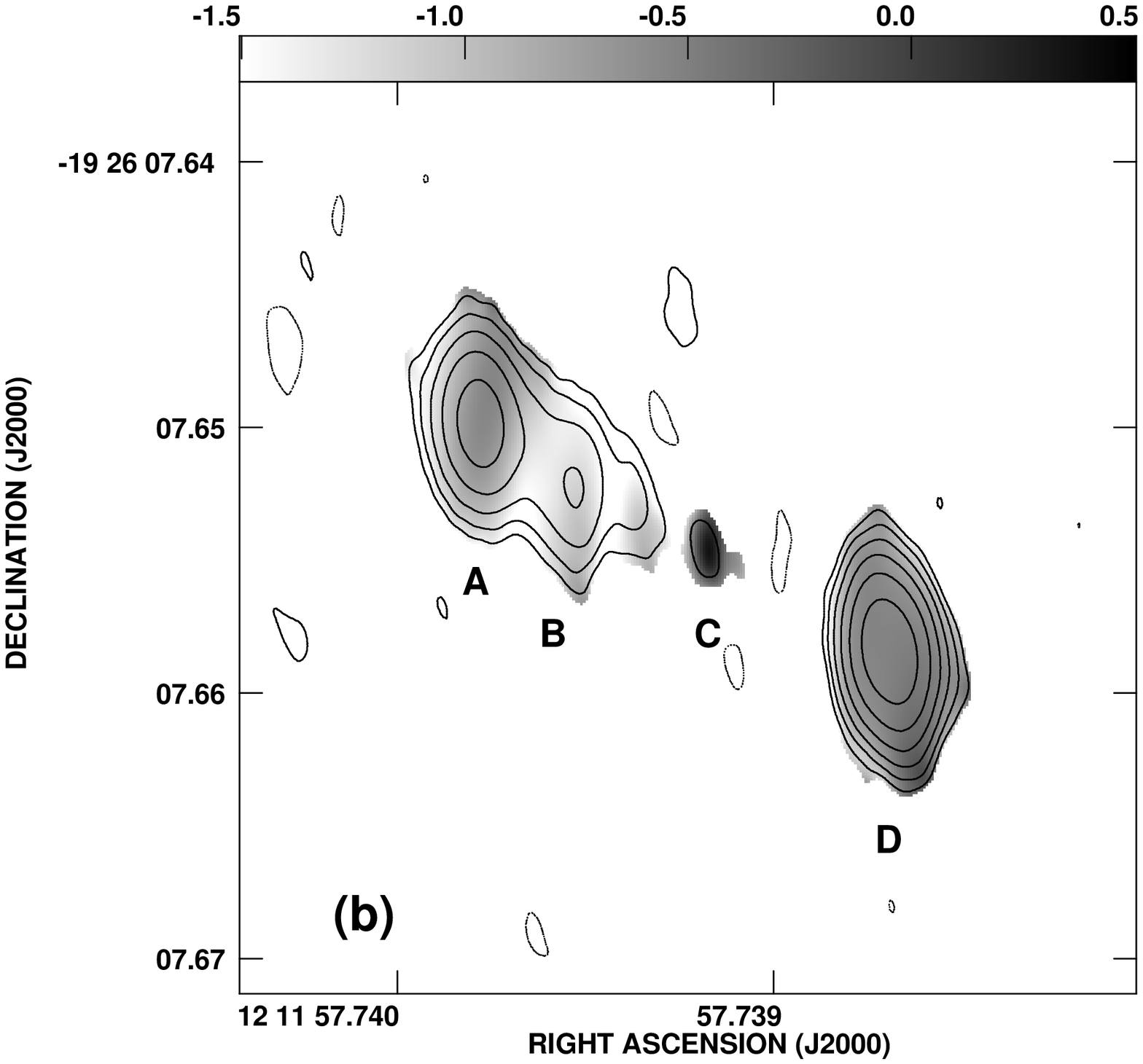}
\caption{{\bf (a)} Total intensity contours of J1211{\tt -}1926 at 5 GHz. 
{\bf (b)} Total intensity
contours from 15 GHz with a greyscale map of the spectral index
between 5 and 15 GHz overlaid.  The greyscale range is from 
$-$1.5 to 0.5.  The restoring beam for both images is 4 $\times$ 1.6 
mas in position angle 0\deg.
\label{fig4}}
\end{figure}
\clearpage

\begin{figure}
\vspace{20.4cm}
\includegraphics{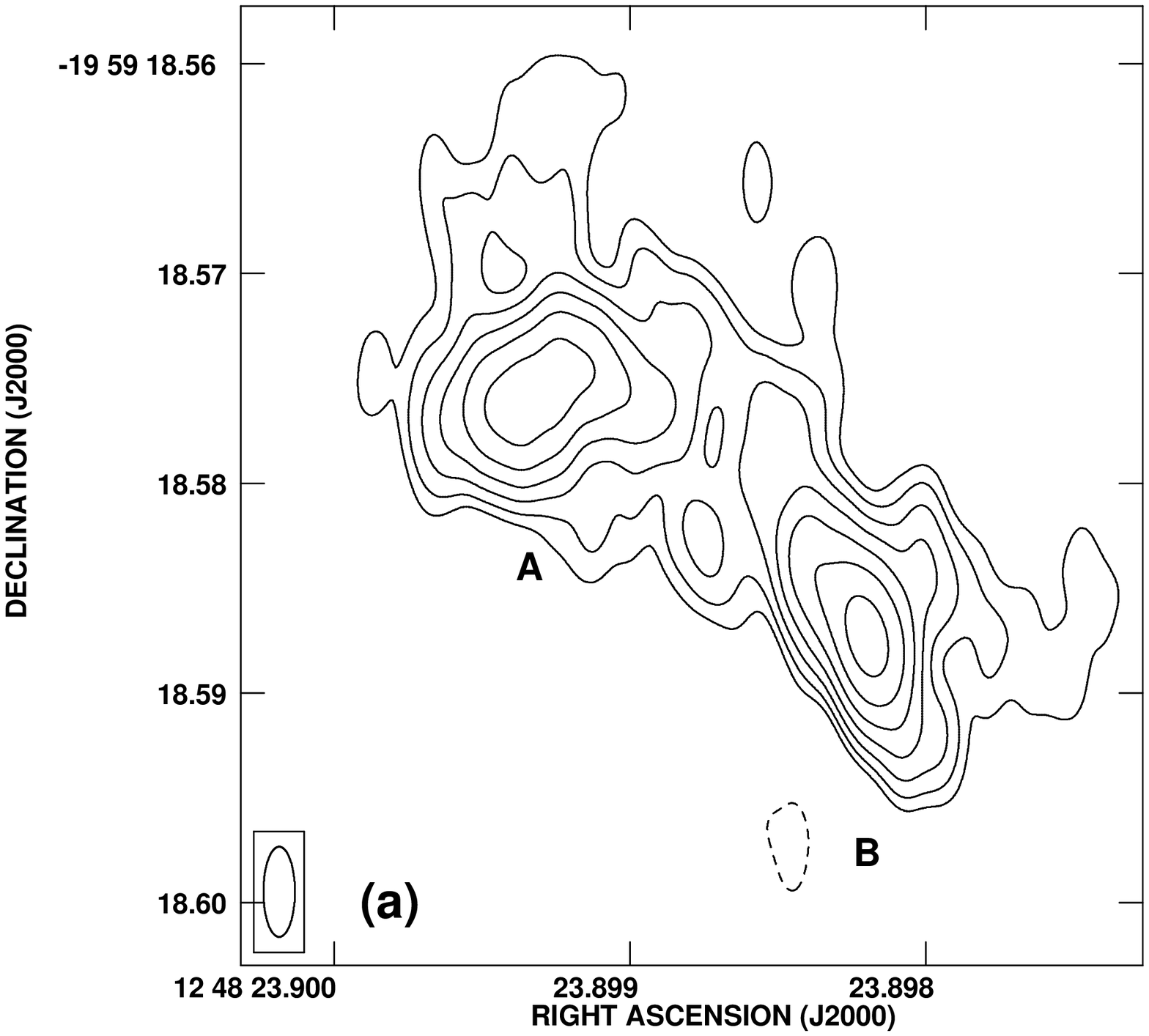}
\includegraphics{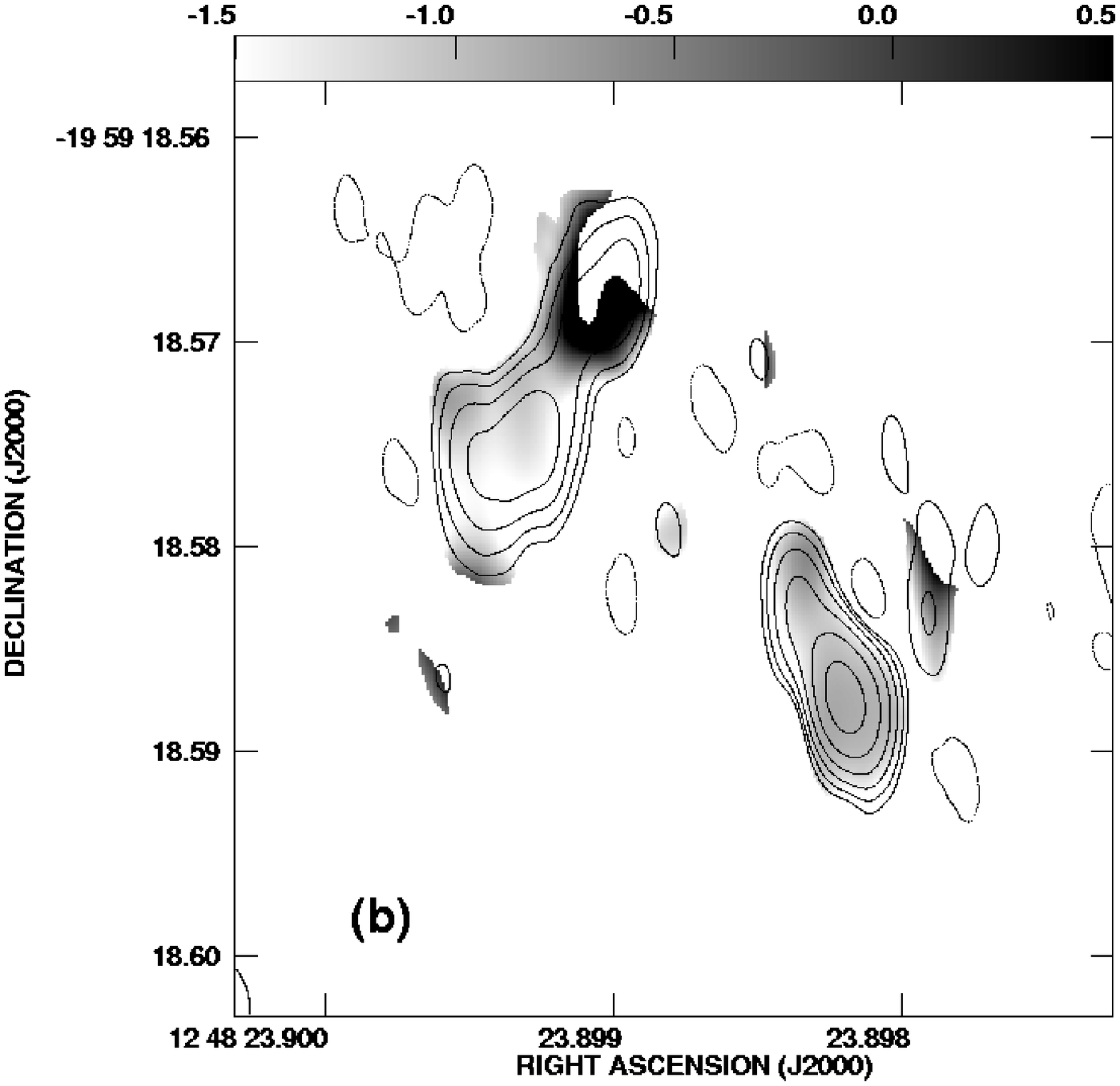}
\caption{{\bf (a)} Total intensity contours of J1248{\tt -}1959 at 5 GHz. 
{\bf (b)} Total intensity
contours from 15 GHz with a greyscale map of the spectral index
between 5 and 15 GHz overlaid.  The greyscale range is from 
$-$1.5 to 0.5.  The restoring beam for both images is 4.3 $\times$ 1.5
mas in position angle 0\deg.
\label{fig5}}
\end{figure}
\clearpage

\begin{figure}
\vspace{20.4cm}
\includegraphics{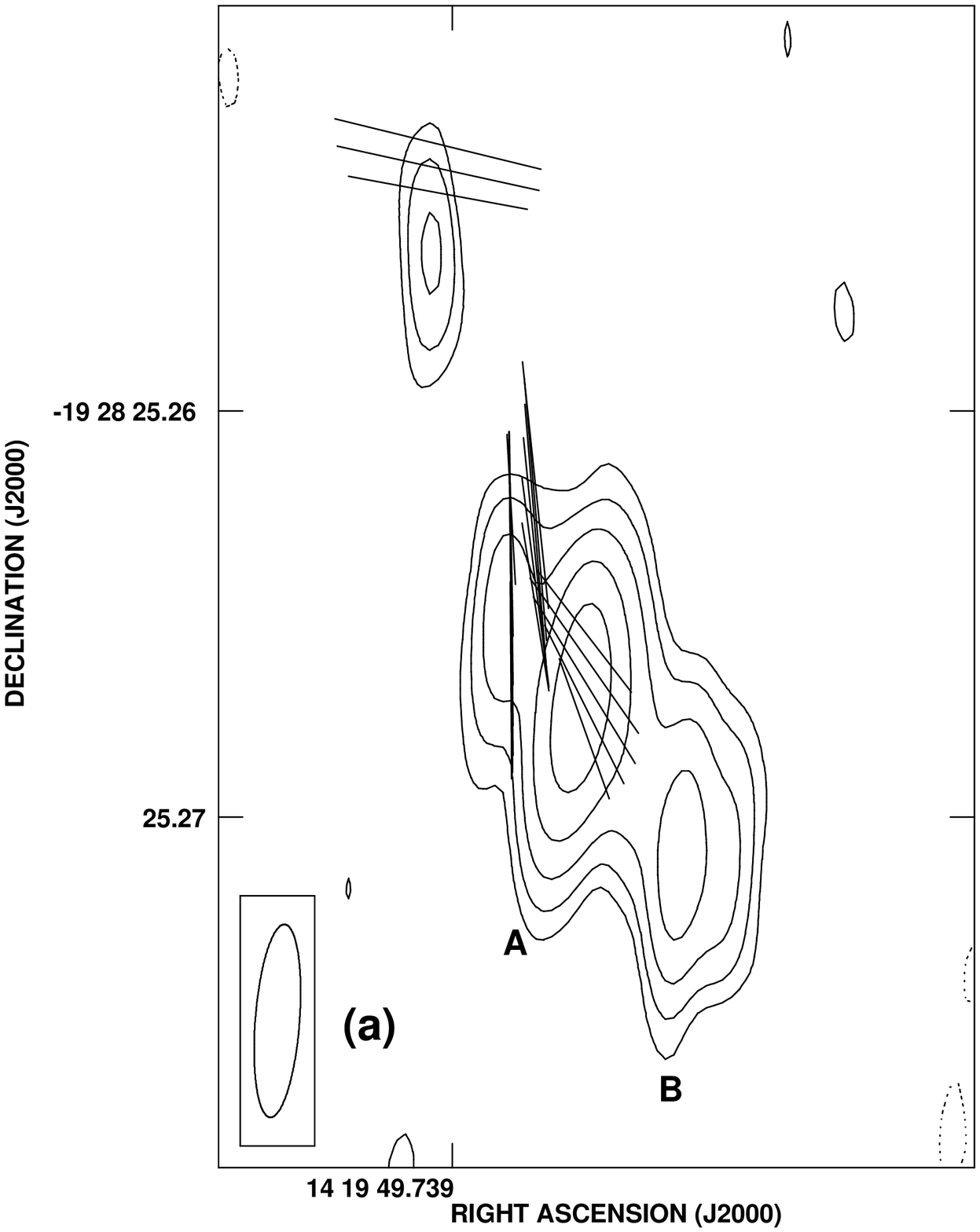}
\includegraphics{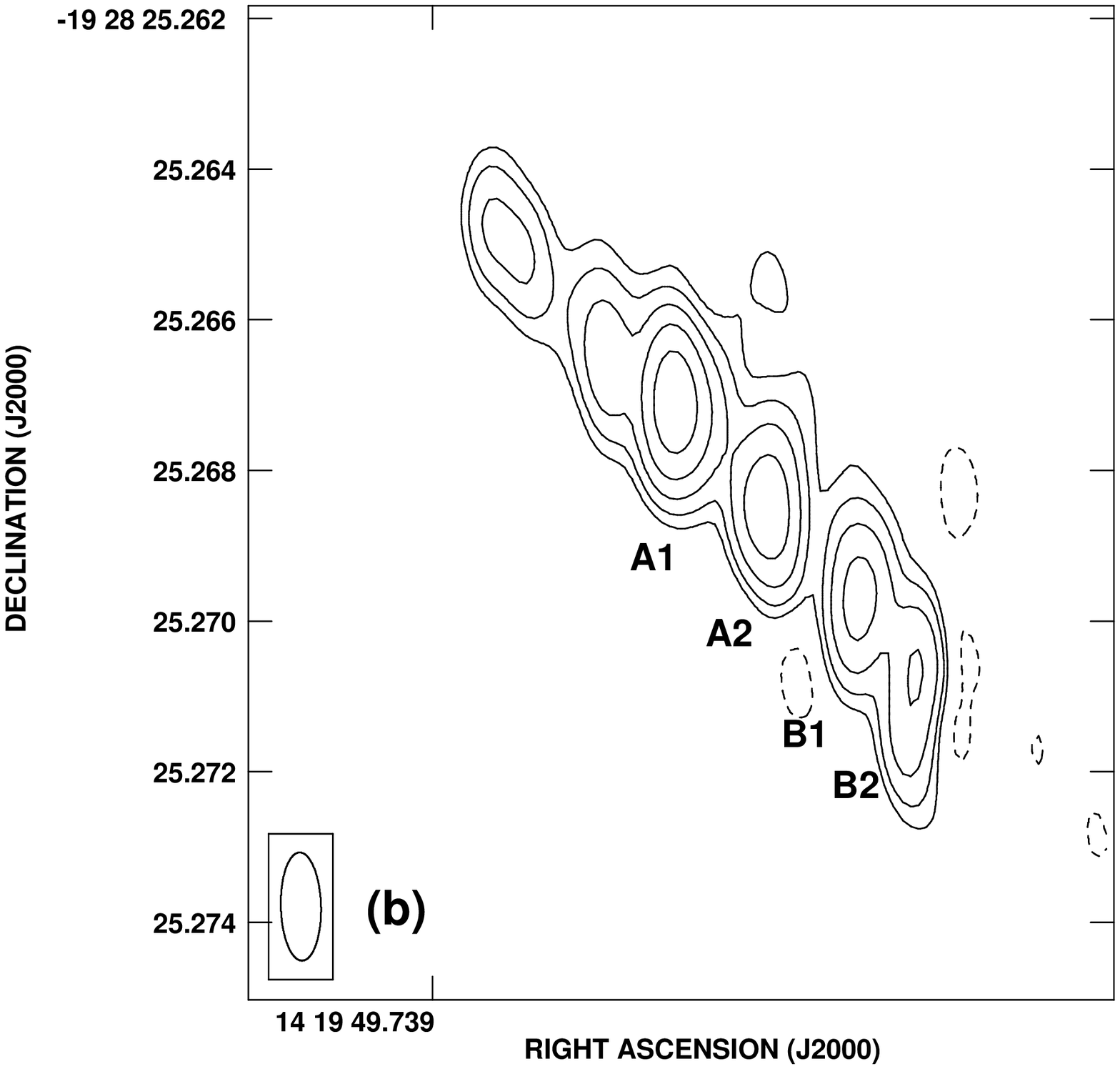}
\caption{{\bf (a)} Total intensity contours of J1419{\tt -}1928 at 5 GHz
with polarization vectors overlaid. 
The restoring beam is 4.8 $\times$ 1.1
mas in position angle $-$4\deg.{\bf (b)} Total intensity
contours from 15 GHz at full resolution.
The restoring beam is 1.44 $\times$ 0.53 mas in position angle 1\deg.
\label{fig6}}
\end{figure}
\clearpage

\begin{figure}
\vspace{20.4cm}
\includegraphics{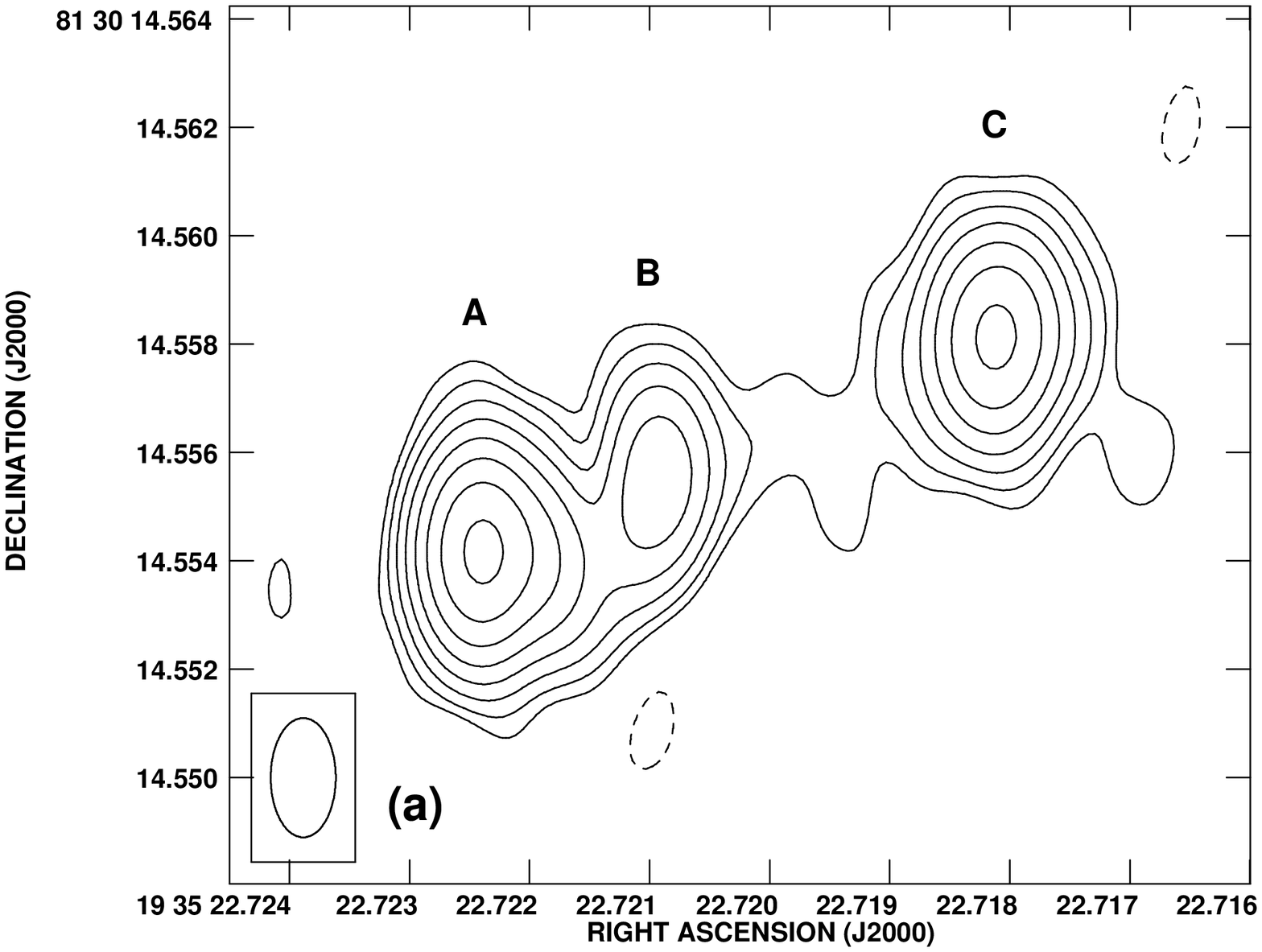}
\includegraphics{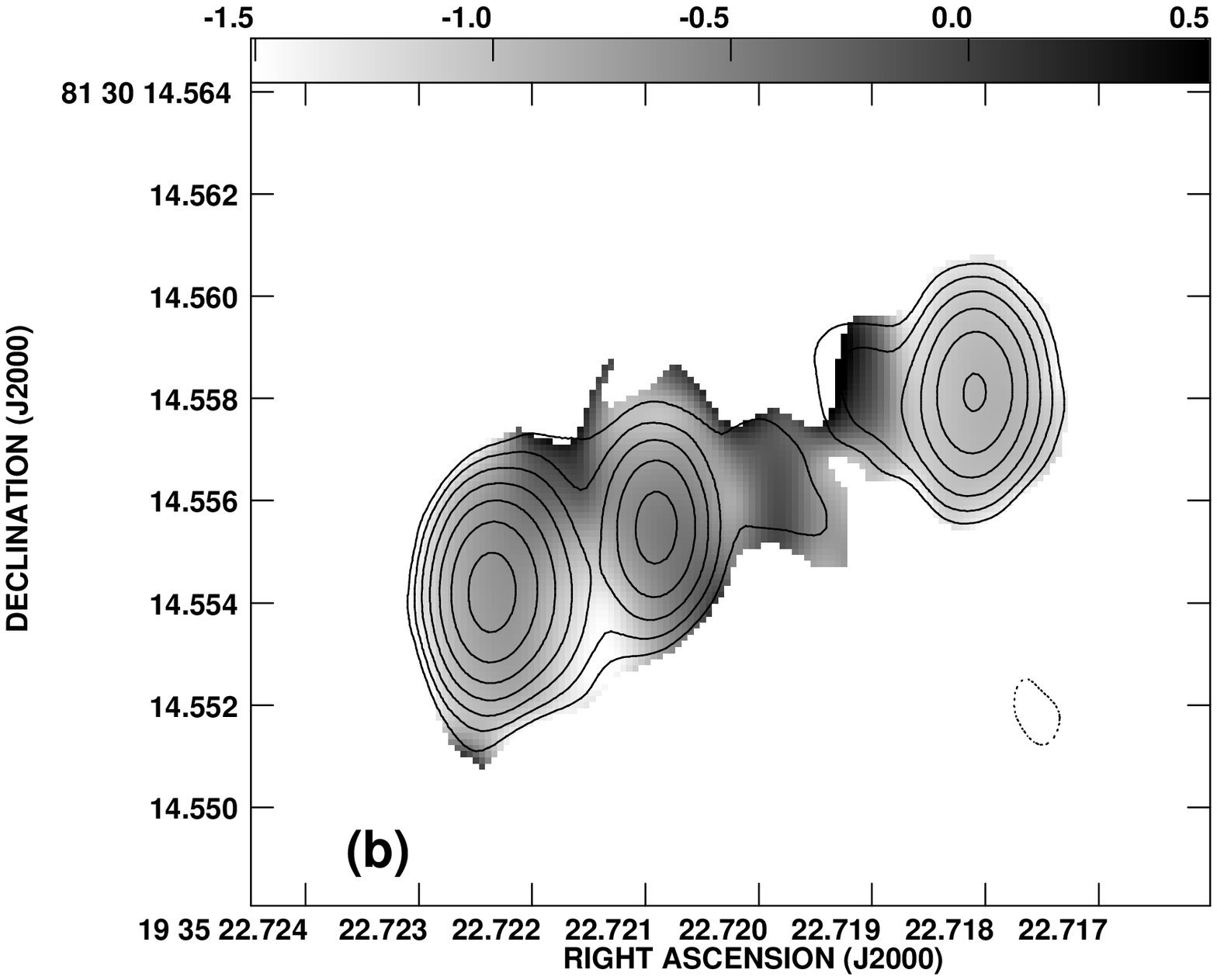}
\caption{{\bf (a)} Total intensity contours of J1935{\tt +}8130 at 5 GHz. 
{\bf (b)} Total intensity
contours from 15 GHz with a greyscale map of the spectral index
between 5 and 15 GHz overlaid.  The greyscale range is from 
$-$1.5 to 0.5.  The restoring beam for both images is 2.2 $\times$ 1.2
mas in position angle 0\deg.
\label{fig7}}
\end{figure}
\clearpage

\begin{figure}
\vspace{20.4cm}
\includegraphics{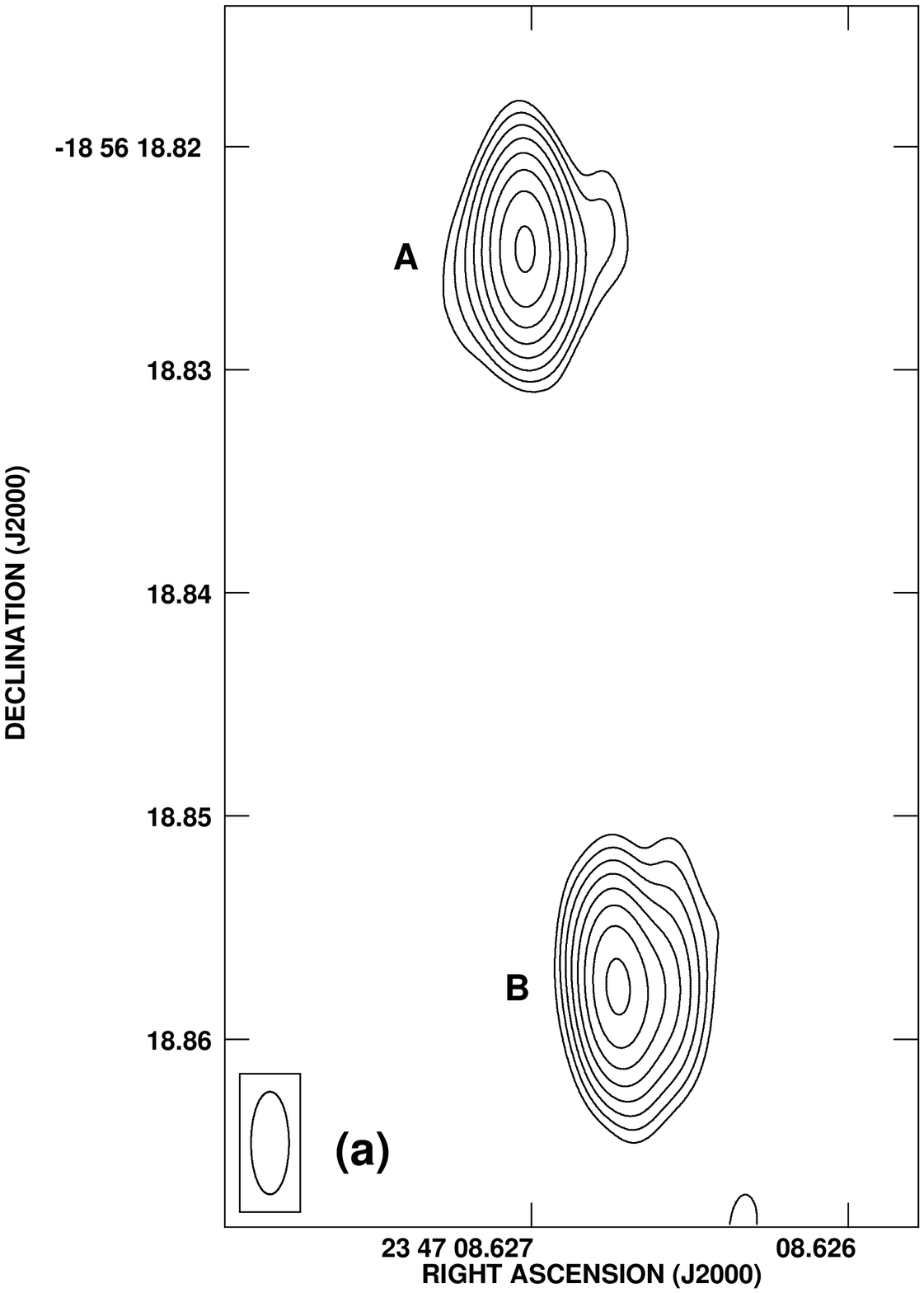}
\includegraphics{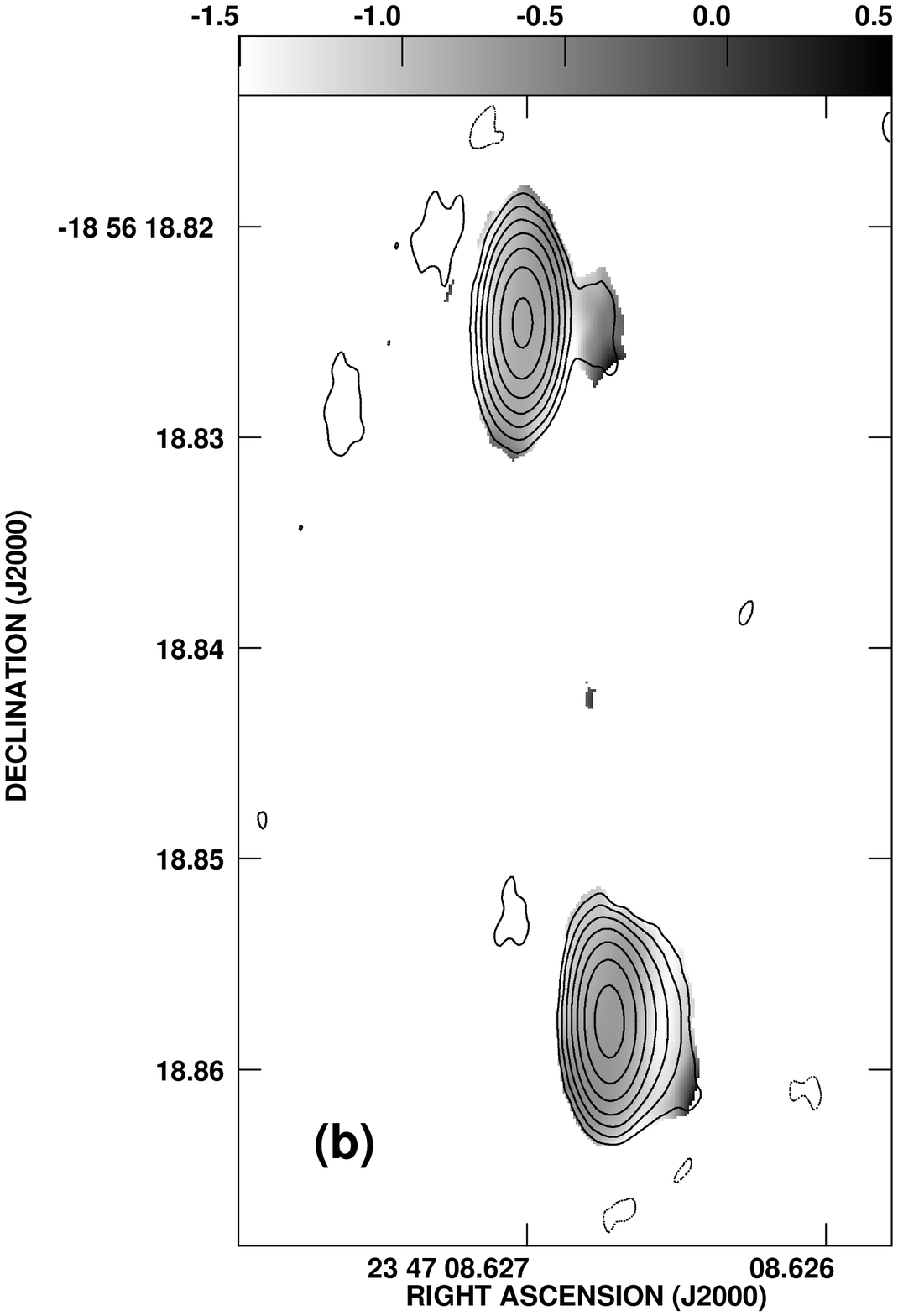}
\caption{{\bf (a)} Total intensity contours of J2347{\tt -}1856 at 5 GHz. 
{\bf (b)} Total intensity
contours from 15 GHz with a greyscale map of the spectral index
between 5 and 15 GHz overlaid.  The greyscale range is from 
$-$1.5 to 0.5.  The restoring beam for both images is 4.6 $\times$ 1.7
mas in position angle 0\deg.
\label{fig8}}
\end{figure}
\clearpage

\renewcommand{\baselinestretch}{1.2}
\scriptsize
\begin{table}[h]
\begin{center}
\caption{CSO Candidates from the VLBA Calibrator Survey Addition\label{tab1}}
~~~\\
\begin{tabular}{llrrrrr}
\tableline\tableline
Source &Alternate& &     \\
Name &Name &{RA} &{Dec} & ID & $M_v$ & $z$  \\
(1) &(2) &(3) &(4)  &(5) & (6) & (7)  \\
\tableline
J0242{\tt -}2132 & OD-267 & 02 42 35.909852 & $-$21 32 25.93498 & G & 17.0 & 0.314 \\
J0425{\tt -}1612 &        & 04 25 53.572659 & $-$16 12 40.24837 & ... & ... & ... \\
J0735{\tt -}1735 & IERS B0733-174 & 07 35 45.812484 & $-$17 35 48.50230 & ... & ... & ... \\
J1211{\tt -}1926 &        & 12 11 57.738693 & $-$19 26 07.65844 & ... & ... & ... \\
J1248{\tt -}1959 & ON-176.2 & 12 48 23.898196 & $-$19 59 18.58763 & Q & 20.5 & 1.275 \\
J1419{\tt -}1928 & CTS 0105 & 14 19 49.738769 & $-$19 28 25.26705 & G & 17.5 & 0.1200 \\
J1935{\tt +}8130 & S5 1939+81 & 19 35 22.722370 & $+$81 30 14.55415 & G & 21.1 & ... \\
J2347{\tt -}1856 & OZ-174 & 23 47 08.626730 & $-$18 56 18.85778 & ... & ... & ... \\
\end{tabular}
\end{center}
Notes -- (1) J2000 source name; (2) alternate source name; (3) Right
Ascension and (4) declination in J2000 coordinates from the VLBA 
Calibrator Survey astrometric determination by Beasley et al. 2001;
(5) Optical host galaxy identification; (6) Optical Magnitude;
(7) redshift (see discussion of individual sources for references).\\
\end{table}

\begin{table}
\begin{center}
\caption{CSO Candidate Image Parameters\label{tab2}}
~~~\\
\begin{tabular}{lrcrrccc}
\tableline\tableline
       &       &       &         &Peak  Flux &  rms & Lowest & \\
Source & Freq. &  Beam & $\theta$ &(mJy &(mJy  & Contour & Status \\
Name   & (GHz) & (mas) & & beam$^{-1}$) &  beam$^{-1}$) & (mJy)
&  \\

\tableline
J0242{\tt -}2132 &5.0 &4.0$\times$1.5&0&330&0.28& 0.83 & CAND \\
                &15.2 &4.0$\times$1.5&0&151&0.49& 1.5 & \\
J0425{\tt -}1612 &5.0 &3.6$\times$1.4&0&57.2 &0.22& 0.7 & CSO\\
                &15.2 &3.6$\times$1.4&0&19.3 &0.25& 0.8 & \\
J0735{\tt -}1735 &5.0 &6.0$\times$1.8&17&538&0.53& 1.6 & CSO \\
                &15.2 &6.0$\times$1.8&17&274&0.32& 1.0 & \\
J1211{\tt -}1926 &5.0 &4.0$\times$1.6&0&85.1 &0.23& 0.7 & CSO \\
                &15.2 &4.0$\times$1.6&0&47.7 &0.29& 0.9 & \\
J1248{\tt -}1959 &5.0 &4.3$\times$1.5&0&245&0.90& 2.7 & CAND \\
                &15.2 &4.3$\times$1.5&0&95.9 &0.69& 2.3 & \\
J1419{\tt -}1928 &5.0 &4.8$\times$1.1&0&94.9&1.30& 3.9 & CAND\\
                &15.2&1.44$\times$0.53&1.4&45.3&0.56& 1.7 & \\
J1935{\tt +}8130 &5.0 &2.2$\times$1.2&0&175&0.38& 1.1 & CAND \\
                &15.2 &2.2$\times$1.2&0&80.3 &0.30& 0.9 & \\
J2347{\tt -}1856 &5.0 &4.6$\times$1.7&0&176&0.29& 0.9 & CSO \\
                &15.2 &4.6$\times$1.7&0&80.1 &0.29& 0.9 & \\
\tableline
\end{tabular}
\end{center}
\tablenum{2}
\end{table}
\clearpage

\renewcommand{\baselinestretch}{1.2}
\scriptsize
\begin{table}
\tablenum{3}
\begin{center}
\caption{CSO Candidate Component Modelfits and Polarization\label{tab3}}
~~~\\
\begin{tabular}{lcrrrrrrr}
\tableline\tableline
Source &Comp. &$S_5$ &$S_{15}$ &$\alpha^{5}_{15}$&P$_{5}$ & m$_{5}$ \\
 (1) & (2)  & (3)  &(4) &(5) &(6) &(7)\\
\tableline
J0242{\tt -}2132 &A& 133.3 & 100.6 & $-$0.25 & $<$0.89 & $<$0.055 \\
                 &B& 358.9 & 168.7 & $-$0.66 & 1.32 & 0.004 \\
                 &C& 190.0 &  59.2 & $-$1.02 & $<$0.89 & $<$0.026 \\

J0425{\tt -}1612 &A& 50.8 & 29.0 & $-$0.49 & $<$0.72 & $<$0.113 \\
                 &B& 83.2 & 38.7 & $-$0.67 & $<$0.72 & $<$0.019 \\
                 &C& 4.2  & 5.3 & 0.20 & $<$0.72 & $<$0.350 \\
                 &D& 107.1 & 34.0 & $-$1.00 & $<$0.72 & $<$0.013 \\

J0735{\tt -}1735 &A& 673.4 & 340.1 & $-$0.60 & $<$1.21 & $<$0.042 \\
                 &B& 300.2 & 161.4 & $-$0.54 & 1.10 & 0.002 \\
                 &C& 234.2 &  62.2 & $-$1.16 & $<$1.21 & $<$0.012 \\

J1211{\tt -}1926 &A& 39.4 &  20.4 & $-$0.59 &  $<$0.83 & $<$0.021 \\
                 &B& 28.2 &   8.6 & $-$1.06  & $<$0.83 & $<$0.030 \\
                 &C& 1.0  &  1.3 & 0.23  & ... & ... \\
                 &D& 135.1 & 73.9 & $-$0.53  & $<$0.83 & $<$0.009 \\

J1248{\tt -}1959 &A& 712.4 & 248.7 & $-$0.92 & $<$0.91 & $<$0.021 \\
                 &B& 465.9 & 200.3 & $-$0.74 & $<$0.91 & $<$0.021 \\

J1419{\tt -}1928 &A& 164.1 & 121.1 & $-$0.27 & 2.28 & $<$0.025 \\
                 &B& 83.2  &  55.2 & $-$0.36 & $<$2.16 & $<$0.048 \\

J1935{\tt +}8130 &A& 213.2 & 99.9 & $-$0.66 & $<$1.13 & $<$0.006 \\
                 &B& 76.2  & 31.9 & $-$0.76 & $<$1.13 & $<$0.034 \\
                 &C& 118.9 & 40.7 & $-$0.94 & $<$1.13 & $<$0.013 \\

J2347{\tt -}1856 &A& 213.1 & 80.2  & $-$0.85 & $<$1.09 & $<$0.007 \\ 
                 &B& 247.4 & 104.4 & $-$0.75 & $<$1.09 & $<$0.006 \\

\end{tabular}
\end{center}
Notes -- (1) J2000 source name; (2) component (see individual figures); 
(3) integrated flux density of the component at 4.98 GHz as derived from 
Gaussian model-fitting to the visibility data; 
(4) integrated flux density of the component at 15.2 GHz as derived from 
Gaussian modelfitting to the visibility data; (5) spectral index 
between 4.98 and 15.2 GHz; (6) peak polarized intensity 
of the component at 5 GHz, or 
3$\sigma$ limit; (7) fractional polarization at the peak or 3$\sigma$ limit.
\\
\tablenum{3}
\end{table}
\clearpage

\end{document}